# Magnetically Controllable Multimode Interference in Topological Photonic Crystals

Weiyuan Tang[1#], Mudi Wang[2#], Shaojie Ma[1,3], C.T. Chan[2*], Shuang Zhang[1*]

[1]Department of Physics, The University of Hong Kong, Hong Kong, China

[2]Department of Physics, The Hong Kong University of Science and Technology, Hong Kong, China

[3]Department of Optical Science and Engineering, Fudan University, 200433, Shanghai, China

[#]these authors contributed equally to this work

[*]e-mail: phchan@ust.hk; shuzhang@hku.hk

Topological photonic insulators show promise for applications in compact integrated photonic circuits due to their ability to transport light robustly through sharp bendings. The number of topological edge states relies on the difference between the bulk Chern numbers across the boundary, as dictated by the bulk edge correspondence. The interference among multiple topological edge modes in topological photonics systems may allow for controllable functionalities that are particularly desirable for constructing reconfigurable photonic devices. In this work, we demonstrate magnetically controllable multimode interference based on gyromagnetic topological photonic insulators that support two unidirectional edge modes with different dispersions. We successfully achieve controllable power splitting in experiments by engineering multimode interference with the magnetic field intensity or the frequency of wave. Our work demonstrates that manipulating the interference among multiple chiral edge modes can facilitate the advancement of highly efficient and adaptable photonic devices.

Over the last decade, topological photonics[1–32] has emerged as a promising field of research, drawing increasing attention for its intriguing physics and potential applications. Inspired by the quantum Hall effects and quantum spin Hall effects in condensed matter physics, significant research efforts have been devoted to investigating and achieving analogous topological phases and band theory in photonic systems. Among the interesting effects associated with topological systems, one of most remarkable is the presence of non-reciprocal topological edge modes, which arise from the broken time-reversal symmetry. The number of the chiral edge modes (CEMs) is determined by the bulk topological invariants according to bulk-edge correspondence[33], and they are robust to general types of disorders. Multiple CEMs can arise when the Chern number difference across the interface is greater than one.

Conventionally, multimode interference is realized by modifying the size of waveguides[34–38], including their width or length, or controlling the refractive indices of the system with electrooptic effect[39,40]. However, back reflection loss[41] is a key problem that leads to performance limitations of traditional multimode interference devices. Based on the backscattering-immune property of topologically protected CEMs, topological multimode waveguide can overcome this limitation with one-way propagation, and support higher mode density and coupling efficiency[18], thereby offer opportunities for design of novel topological devices for power manipulation.

To date, most proposed solutions for manipulating multimode interference in topological systems involve modifying the structural dimensions of channels[18,42–48], such as, by changing the length of the incident channel to adjust the propagation phase of the incident waves[18]. It has also been reported that adjustable current partition of valley edge states can be achieved by controlling the coupling strength between the incident and outgoing channels with bending angles[45–47]. However, the requirement for varying the geometry of the system with these techniques severely restricts their practical utility. Recently, some researchers have proposed to manipulate interference of multiple CEMs with structure-independent parameters, such as external magnetic field[49–51] and frequency[50,52]. However, experimental demonstrations of multimode interference in topological photonic crystals are still lacking. Though the inverted, doublefold and direct images of input field can be located by beam length[34,51], the underlying mechanisms of the process of multimode interference still remain elusive, and further investigation is required to elucidate the underlying principles.

In this work, we achieve controllable multimode interference in a topological photonic heterostructure waveguide formed by two gyromagnetic photonic crystals, where the oppositely biased magnetic fields broke the time-reversal symmetry. The power splitting ratio originating from multimode interference is controlled by the propagating phase difference between two CEMs (Fig.1a), which depends on the incident frequency (Fig.1b) or magnetic field intensity (Fig.1c). Using the transfer matrix method, we develop a theoretical model to elucidate the fundamental mechanisms underlying tunable multimode interference in the heterostructure. Our experimental results confirm the controllability of multimode interference of CEMs through magnetic field or frequency manipulation.

**Result**

**The mechanism of multimode interference in Topological Photonic Crystals.** Fig. 2a depicts an H-shaped heterostructure waveguide with two domains, A and B, denoted in green and blue, respectively. Each domain is composed of a triangular lattice of yttrium iron garnet (YIG) rods with a lattice constant $a = 16$ mm and a rod radius $r = 2$ mm. The YIG rods are magnetized by an external magnetic field applied in the positive (negative) vertical direction to domain A (B). The domains are separated by an air gap with a width of $w = 1.09a$, which modifies the dispersion behavior of CEMs. We focus on the transverse magnetic (TM) modes for the formation of the bulk bands of domains A and B, as shown in Fig. 2b. The Dirac points in the bulk band structure of domains A and B is gapped due to the breaking of time-reversal symmetry in the presence of external magnetic field, while the Chern numbers of the relevant bulk bands are indicated in the panel. Two gapless unidirectional edge states are present in the air gap (Fig. 2c), as determined by the difference between the gap Chern numbers of domain A and B through the bulk-edge correspondence[33]. Domains A and B each contributes one edge mode and these two topological modes hybridize to form mode 1 and mode 2 that propagate in the air gap, with quasi-asymmetric and quasi-symmetric distribution of electric field $E_z$ about $y = 0$, respectively, as shown in the right pane of Fig. 2c.

Based on the backscattering-immune property of topologically protected CEMs at the intersections of this H-shaped waveguide (S and P point), we then develop a theoretical model that captures the essence of the propagation properties of CEMs in our system with the transfer-matrix method:

$$\begin{bmatrix} B_1 \\ B_2 \end{bmatrix} = \begin{bmatrix} \frac{1}{\sqrt{2}} & \frac{1}{\sqrt{2}} \\ -\frac{1}{\sqrt{2}} & \frac{1}{\sqrt{2}} \end{bmatrix} \begin{bmatrix} e^{ik_1 l} & 0 \\ 0 & e^{ik_2 l} \end{bmatrix} \begin{bmatrix} \frac{1}{\sqrt{2}} & -\frac{1}{\sqrt{2}} \\ \frac{1}{\sqrt{2}} & \frac{1}{\sqrt{2}} \end{bmatrix} \begin{bmatrix} A_1 \\ A_2 \end{bmatrix} \quad (1)$$

where $A_1(A_2)$ represents the CEM excited in channel S1 (S2), while $B_1(B_2)$ denotes the mode probed in channel P1 (P2). The source/detection points are located far from the connection point (S or P point) of the H-shaped waveguide. The wave numbers $k_1$ and $k_2$ represent the wave numbers of mode 1 and mode 2, respectively, and the length of the channel M is denoted by $l$ (see details in the Supplementary Note 1). To visually demonstrate the interference between two CEMs, we here introduce the power splitting ratio of channel P1 and P2, which is defined as:

$$R = \frac{|B_1|^2}{|B_1|^2 + |B_2|^2} \quad (2)$$

The power splitting ratios for the source being placed in channel S1 and S2 can be obtained from Eq. (1) and Eq. (2) as:

$$R_{S1} = \frac{|e^{i\Delta\varphi}+1|^2}{|e^{i\Delta\varphi}+1|^2 + |e^{i\Delta\varphi}-1|^2} \quad (3)$$

$$R_{S2} = \frac{|e^{i\Delta\varphi}-1|^2}{|e^{i\Delta\varphi}+1|^2 + |e^{i\Delta\varphi}-1|^2} \quad (4)$$

where $\Delta\varphi = (k_2 - k_1)l$ is the phase difference between the mode 1 and mode 2 propagating in the same direction. Remarkably, the power splitting ratio is regulated by the interference between two CEMs, which is solely subject to the phase difference $\Delta\varphi$ within the framework of transfer matrix description. To further unveil the mechanisms of controllable multimode interference, the dependence of propagating phase difference

Δ$\varphi$ on the frequency and magnetic field are revealed in Fig. 2d-e. These figures show that the phase difference Δ$\varphi$ increases monotonically with frequency for a given magnetic field but decreases monotonically with increasing magnetic field intensity at a fixed frequency. Consequently, the interference between two CEMs in our heterostructure waveguide is customizable by adjusting magnetic field intensity or incident frequency, leading to controllable power splitting ratio.

**Experimental demonstration of controllable multimode interference.** Figure 3a shows a snapshot of the experimental setup. In the photonic heterostructure, the YIG rods are sandwiched between two flat aluminum plates. Several small holes are drilled for probes to access the field through the top plate. Note that both the holes diameter and slots width are 2 mm, which have a negligible effect on the wave propagation within the heterostructure. The upper and lower boundaries of the heterostructure are interfaced with absorbent materials, while the remaining boundaries are interfaced with aluminum blocks. Both the aluminum plates and blocks imitate the PEC boundaries to prevent the transverse magnetic modes from escaping. We start by verifying the unidirectional propagation characteristic of the CEMs. A point source is sequentially placed at the two ends of the heterostructure waveguide (point S and point P in Fig. 2a) and the transmission is measured at the opposite end. The resulting measurement is shown in Fig. 3b, where S21 and S12 represent the measured transmission spectra from S to P and P to S, respectively. The asymmetric transmission in the bandgap explicitly validates the chirality of the CEMs. To obtain the dispersions of the CEMs, we place a needle source at different location to excite each CEM at its maximum electric field strength. Then we measure the relevant out-of-plane electric field distribution $E_z$ via near-field scanning. After applying Fourier transformation, the dispersions of the mode 1 and mode 2 are plotted using color maps in Fig. 3c-d, respectively, which agree well with the numerical results, despite the slight horizontal lines resulting from Fabry–Pérot effects caused by the scattering loss of the experimental setup.

To achieve controllable interference of CEMs with a magnetic field, one can tune the magnetic field strength by adjusting the distance between the YIG rod and the magnet pillar. The correlation between the magnetic field and the distance is measured by a magnetometer, with the results presented in Fig. S3. As expected, the magnetic field strength $B$ decreases monotonically with distance h, enabling us to modify the magnetic field applied to the YIG rods. We probe the transmission at channel P1 and P2 when the excitation source is respectively placed at channel S1 and S2 (denoted by stars in Fig.2a) with an incident frequency $f = 12.13$ GHz. In Fig. 4a, the measured power splitting ratio is represented by circles, which show good agreements with the numerical results plotted by solid lines. Fig. 4b shows that the plot of the theoretically calculated power splitting ratio is in good agreement with the numerical result as a function of magnetic field strength, indicating the validity of the transfer-matrix method.

The energy distribution along the P1 and P2 channels is further investigated under three different magnetic fields, represented by the grey dashed lines in Fig. 4a. For clarity, we focus on the configuration in which channel S1 is excited. At $B = 0.055$ T, the power splitting ratio is roughly 0.5 and the power flow is nearly symmetric about $y = 0$ (Fig. 4c), which indicates that the CEMs divided equally. As the magnetic field intensity increases to $B = 0.097$ T, the CEMs are primarily transferring energy into channel P1 but strongly suppressed in channel P2, reaching a power splitting ratio close to 100%, as shown in Fig. 4d. With further increase of magnetic field, the power splitting ratio begins to decline. At $B = 0.175$ T, as shown by Fig. 4e, most of the power flows into channel P2. The measured signals attenuate along the sample edges mainly due to the scattering loss caused by small air gaps between the YIG rods and the aluminum plates as well as the slot drilled on the top aluminum plate for measurement, nonetheless, good agreement between the measured electric field distributions (left panel) and simulated results (right panel) can still be achieved as shown in Fig. 4c-e.

The manipulation of the interference between CEMs by frequency is also demonstrated experimentally. Fig. 5a shows the dependence of the measured splitting ratio of two CEMs at the left intersection on the incident frequency at a fixed magnetic field $B = 0.140$ T, which is generally consistent with the numerical and analytical results as depicted in Fig. 5b. Note that there are some dips resulting from Fabry–Pérot effects introduced by the scattering loss in the measured splitting ratios, due to the inevitable fabrication errors in the wideness of aluminum plates and height of YIG rods. The numerical energy distributions at selected frequencies of $f = 11.90 \text{ GHz}$, $f = 12.04 \text{ GHz}$ and $f = 12.22 \text{ GHz}$ are displayed in Fig. 5c-e, respectively. The power splitting ratios are respectively 0%, 50% and 100% at these three frequencies, which are consistent with the results shown in Fig. 5b. The measured energy distributions, associated with the numerical results, are also shown in the left panel of Fig.5c-e, which is in good agreement with the results shown in Fig. 5a.

**Discussion**

To summarize, we have demonstrated controllable interference of CEMs in a heterostructure photonic waveguide by adjusting the magnetic field and frequency. We have also developed a theoretical model using the transfer-matrix method to elucidate the underling mechanisms of the magnetically tunable multimode interference. The power splitting ratio of our waveguide system within the investigated magnetic field intensity (frequency) range, resulting from the multimode interference, is quantitatively controlled by the propagating phase difference between two CEMs, which is subject to magnetic field intensity and wave frequency. Hence the interference of CEMs at the topological channel intersection can be manipulated without the requirement for modifying the geometric configuration of the waveguide. Our approach provides a panoramic view of the underlying multimode interference. Meanwhile, the realization of controllable multimode interference also sheds light on the related applications in microwave devices such as switches, signal processors, and isolation devices.

**Acknowledgments**

This work was supported by the Research Grants Council of Hong Kong (AoE/P-502/20, 17309021, 16303119) and the Croucher Foundation (CAS20SC01).

**Author contributions**

S. Z. initiated the project, W. T. designed the configuration, conducted numerical simulations and analytical calculations, M. W. and W. T. performed the experiment, W. T., M. W., S. M., C. C., and S. Z. participated in the analysis of the results, C. C. and S. Z. supervised the project. W. T. and M. W. contributed equally to this work.

**Competing interests**

The authors declare no conflict of interest.

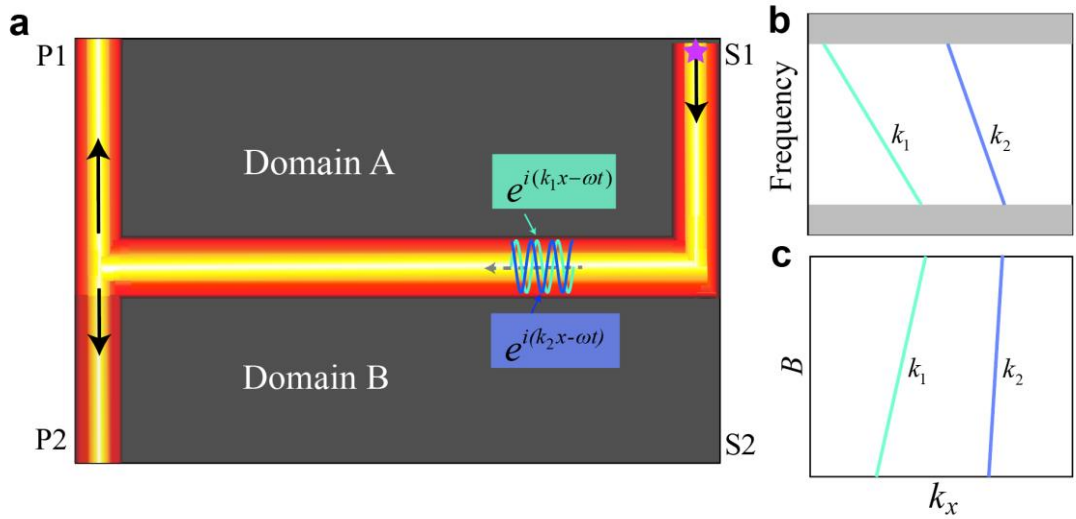

**Fig. 1 The proposed scheme for controllable power splitting through manipulating multimode interference. a** The schematic of a H-shaped heterostructure waveguide composed of two domains, A and B. The star indicates the position of source. **b-c** The differences between the wave number of two CEMs depend on incident frequency (**b**) and the external magnetic field intensity (**c**).

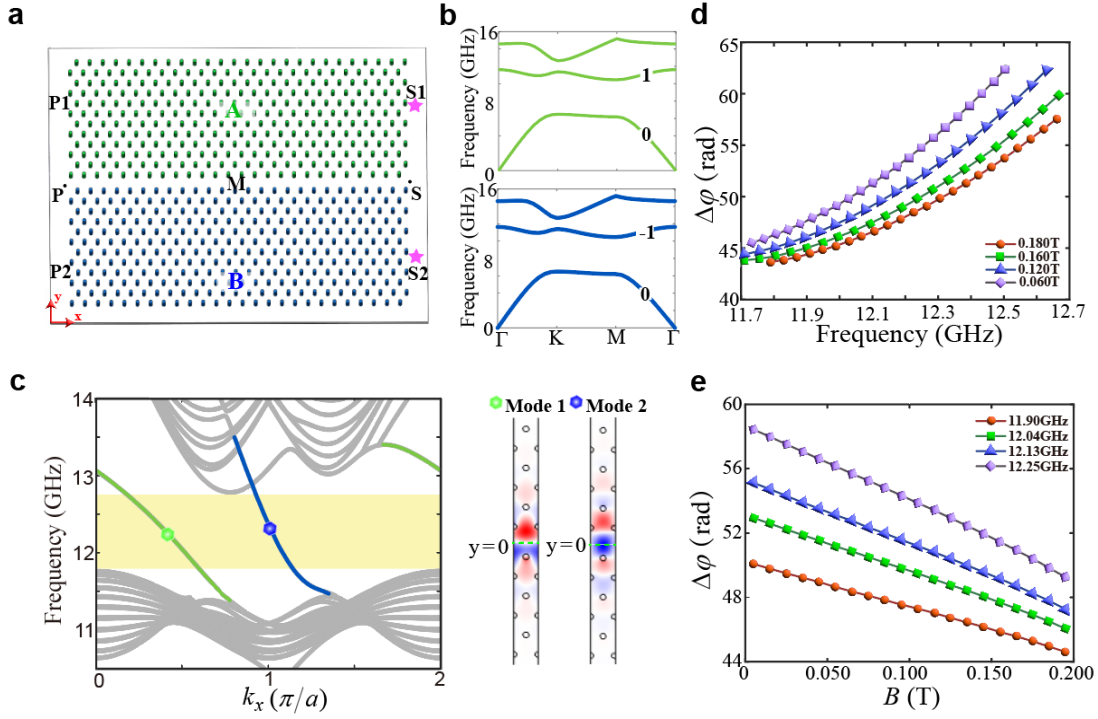

**Fig. 2 The configuration of heterostructure and the phase difference of the two CEMs. a** The schematic of the A|Air|B heterostructure structure, where domain A (B) is subjected to a positive (negative) external magnetic field. The width of the air gap is $w$. **b** The bulk band structures for domains A and B, with the Chern numbers of the first and second bands being tagged, respectively. **c** Left panel: the projected band structure for heterostructure A|Air|B. The yellow area represents the frequency range of the band gap (11.80-12.73 GHz). The green (blue) line indicates mode 1(2). Right panel: The corresponding Electric field distribution $E_z$. **d-e** The phase difference of two CEMs as functions of magnetic field intensity (**d**) and frequency (**e**).

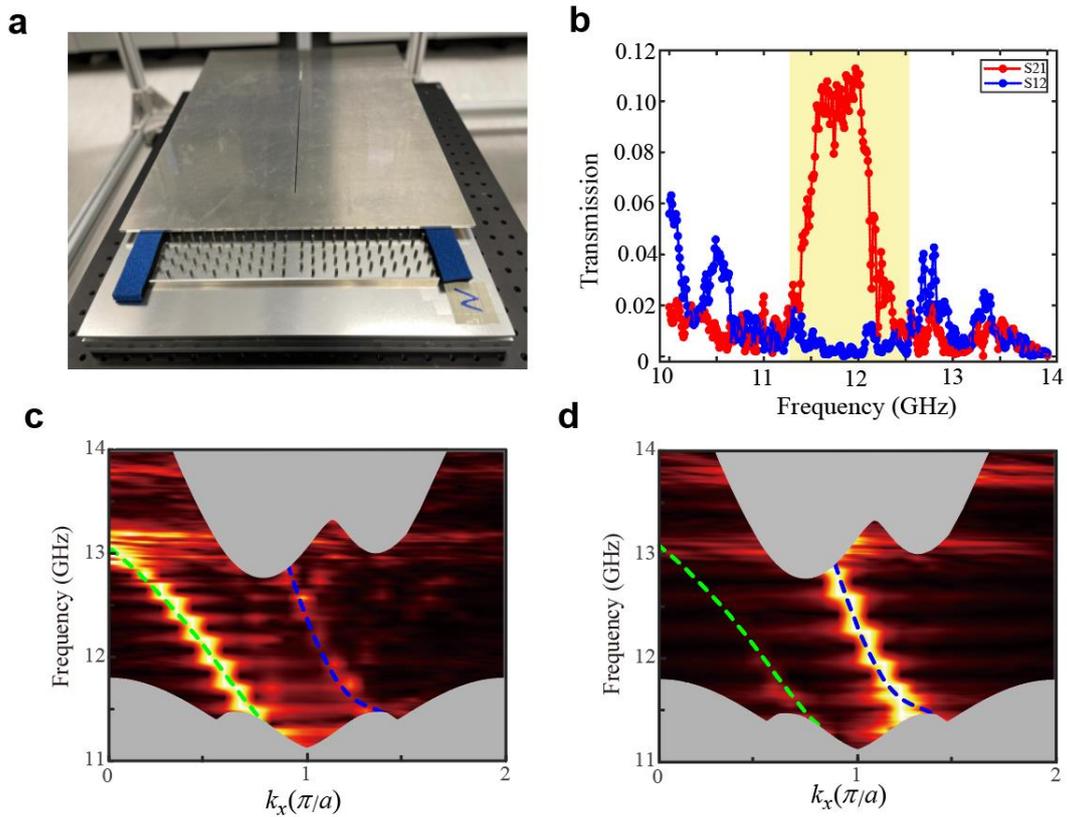

**Fig. 3 Experimental characterization of the properties of CEMs. a** Photograph of the experimental setup. The YIG rods with a height of 6 mm are sandwiched by two aluminum plates to construct a topological photonic crystal waveguide with lattice constant $a = 16$ mm. **b** S21 (S12) is the measured transmission spectrum at point S (P) when the point source is placed on the point P (S) shown in Fig. 2a. **c-d** The selective measurement of mode 1 (**c**) and mode 2 (**d**) at different position of the waveguide. The color map is obtained from experiment data, while the gray area, green and blue dot lines are the numerical results.

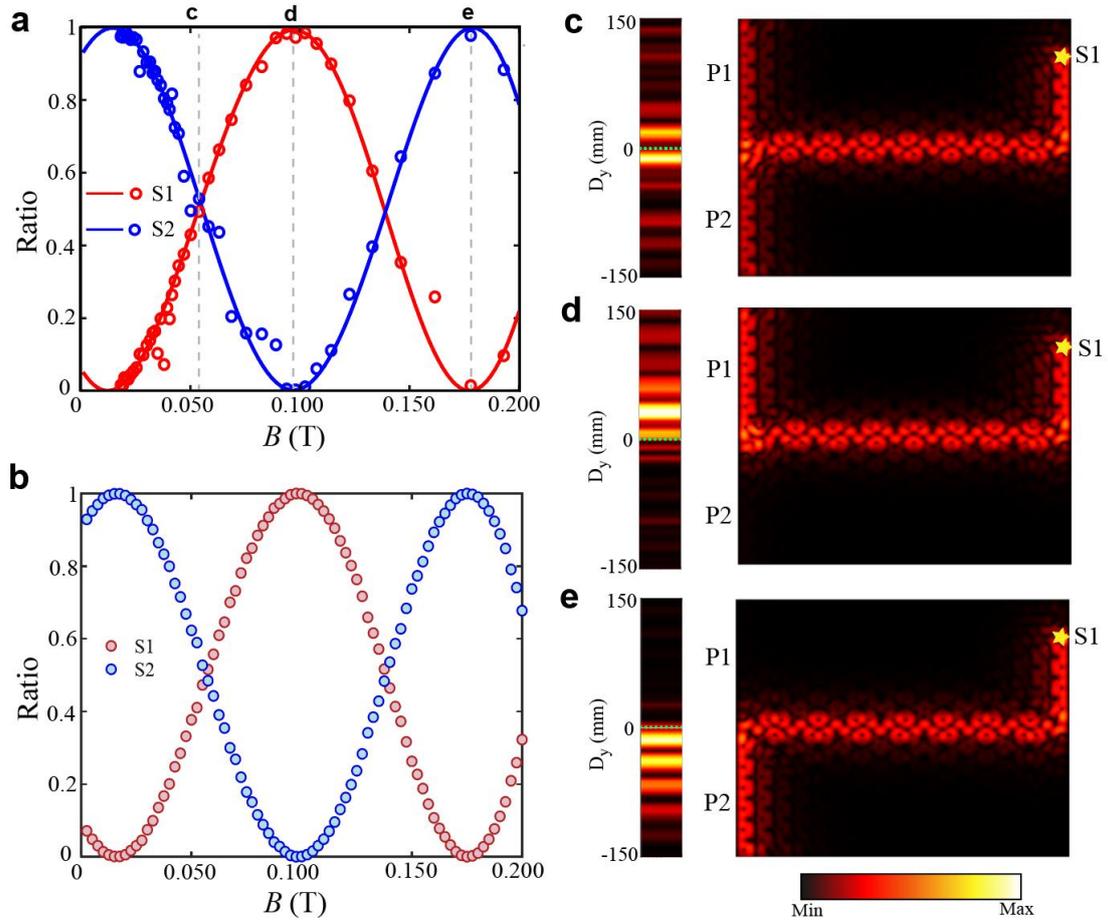

**Fig. 4 Experimental observation of power splitting tuned by changing external magnetic fields. a, b** The measured and theoretical (analytical) power splitting ratio of two CEMs as functions of magnetic field strength. The circles represent experimental data, lines depict the numerical results and the filled circles represent analytical results. The red (blue) color indicates the emitter is placed at channel S1(S2). **c-e** Right panel: numerical electric field distributions at frequency of $f = 12.13$ GHz along the left boundary of the A|Air|B heterostructure under distinct magnetic field strength: $B = 0.055$ T (**c**), $B = 0.097$ T (**d**) and $B = 0.175$ T (**e**), respectively. Left panel: the measured electric field distributions, which are consistent with the associated numerical results.

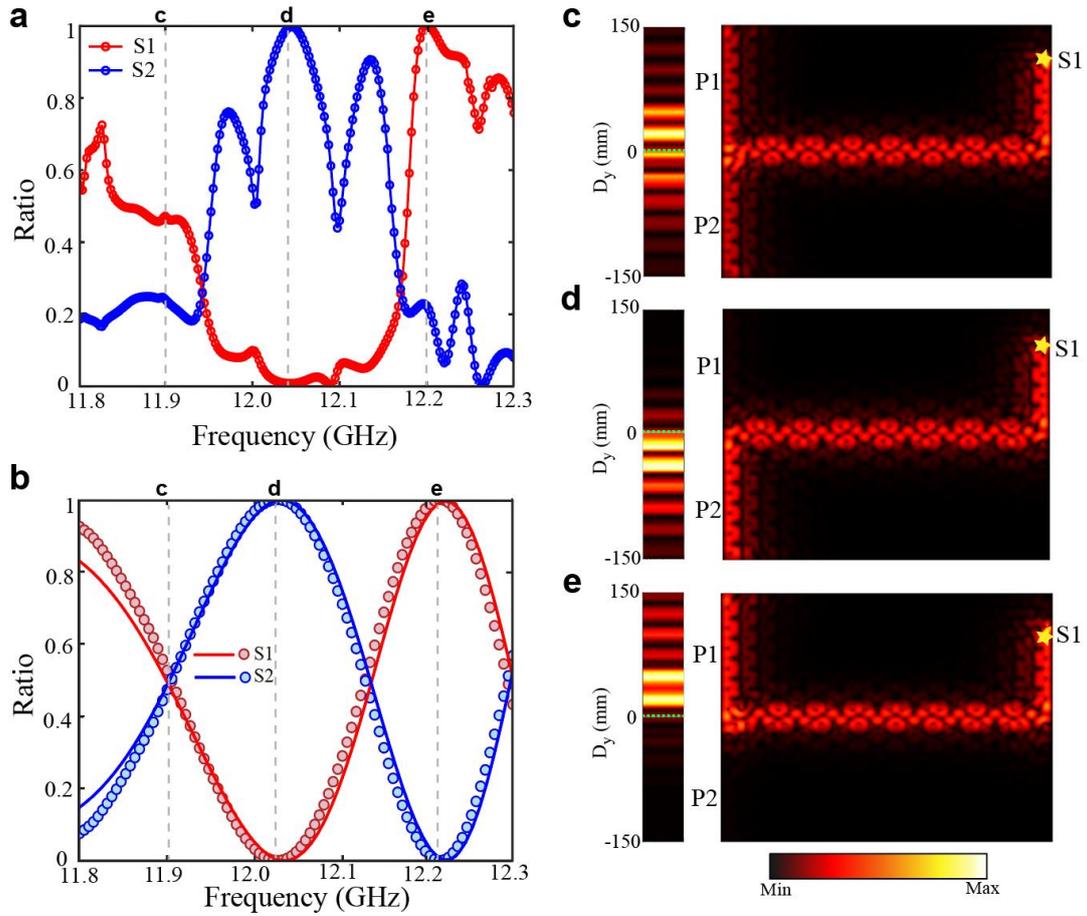

**Fig. 5 Experimental observation of power splitting tuned by different incident frequency. a-b** The measured and analytical power splitting ratio of two CEMs as functions of incident frequency with an external magnetic field of $B = 0.140$ T. The circles represent experimental data, the lines depict the numerical results and the filled circles represent analytical results. The red (blue) color represents the source is excited at channel S1(S2). **c-e** Right panel: numerical electric field distributions along the left boundary of A|Air|B heterostructure with different incident frequencies: $f = 11.90$ GHz (**c**), $f = 12.03$ GHz (**d**) and $f = 12.22$ GHz (**e**), respectively, which are labeled by grey dashed lines in **b**. Left panel: measured electric field distributions for different incident frequencies: $f = 11.90$ GHz (**c**), $f = 12.04$ GHz (**d**) and $f = 12.20$ GHz (**e**), respectively labeled by grey dashed lines in **a**, and is in good agreement with the corresponding numerical results.